\documentclass[12pt,a4paper]{conference}

\usepackage{fancyhdr}
\usepackage{graphicx,amsmath,amssymb,cite}
\usepackage{multind}
\makeindex{author} \makeindex{subject}

\begin{document}

\Chapter{STRUCTURE OF THE SCALARS}
           {Structure of the scalars}{M.~R.~Pennington}

\addcontentsline{toc}{chapter}{{\it M.~.R.~Pennington}} \label{authorStart}

\begin{raggedright}

{\it M.~R.~Pennington\index{author}{Pennington, M.R.}}\\
Institute for Particle Physics Phenomenology\\
Durham University\\
Durham DH1 3LE, U.K.
\bigskip\bigskip

\end{raggedright}

\begin{center}
\textbf{Abstract}
\end{center}
The PDG Tables list more scalar mesons than can fit into one quark model nonet: indeed, even more than can  belong to two multiplets. Consequently, some of these must be states beyond the quark model. So which of these is ${\overline q}q$ or
${\overline{qq}}qq$ or multi-meson molecule or largely glue?
How experiment can help us distinguish between these possibilities is discussed.

\section{Why scalars?}

This meeting has a plenary session and three parallel sessions devoted to the
{\it Scalar Mesons}, and still that is not enough to accommodate all the many contributions to this subject. Others appear 
under the Heavy Flavour label. All these speakers~\cite{MENU2007} have something to say about the \lq\lq Structure of the light scalars''. Here I will address this topic by first explaining why scalars are interesting, for those who do not think about scalars every day of the week. Then I will discuss
when experiment can distinguish between a ${\overline q}q$ meson, a tetraquark state, a ${\overline K}K$ molecule or a glueball. 

So why are the light scalars interesting? This is because they are fundamental. They constitute the Higgs sector of the strong interaction. It is these scalar fields, which have a non-zero vacuum 
expectation value that breaks chiral symmetry and  ensures pions are very light, while giving mass to all other light flavored hadrons~\cite{nambu}. In terms of QCD, quarks  interact  weakly over short distances and they propagate freely. In contrast,
over distances of the order of a fermi, the interactions between quarks, antiquarks and gluons are so strong  that they  generate long range correlations that polarize the vacuum. It is through this medium that quarks have to propagate
across a hadron. The mass of the {\it up} and {\it down} quarks, which is only a few MeV over short distances, becomes 3-400 MeV over the size of a hadron. This change in the behavior of the quark mass is largely produced by a ${\overline q}q$ condensate of the size of $-(240\,{\rm MeV})^3$ for {\it up} and {\it down} quarks. This \lq\lq large'' ${\overline q}q$ condensate  accords with standard Chiral Perturbation Theory~\cite{gasserleut}. Indeed, with its ferromagnetic analogy this supports Nambu's original picture of chiral symmetry breaking
at the hadron level~\cite{nambu}. But then what is the scalar field that is the chiral partner of the pion? In the Nambu-Jona-Lasinio~\cite{njl} and Gell-Mann-Levy~\cite{gml} models it is a single $\sigma$ field. It is the exchange of this particle that was thought to generate the   long range isoscalar force between nucleons. However, if we check the PDG tables~\cite{pdg}, there is not just the $f_0(600)$, which is referred to as the $\sigma$, but a multitude of other isoscalar scalar states: $f_0(980)$, $f_0(1370)$, $f_0(1500)$, $f_0(1710)$, and the more recently discovered $f_0(1830)$~\cite{bes-glue}.  These appear to have isodoublet and isotriplet companions:
$\kappa$, $K_0^*(1430)$, $a_0(980)$ and $a_0(1450)$ below 1.9 GeV.
It is tempting to regard these as the partners of the $f_0$'s, but which?
Different modellers,  many represented at this meeting~\cite{MENU2007}, claim to know the answer.

\begin{figure}[hb]
\begin{center}
\includegraphics[width=6.8  cm]{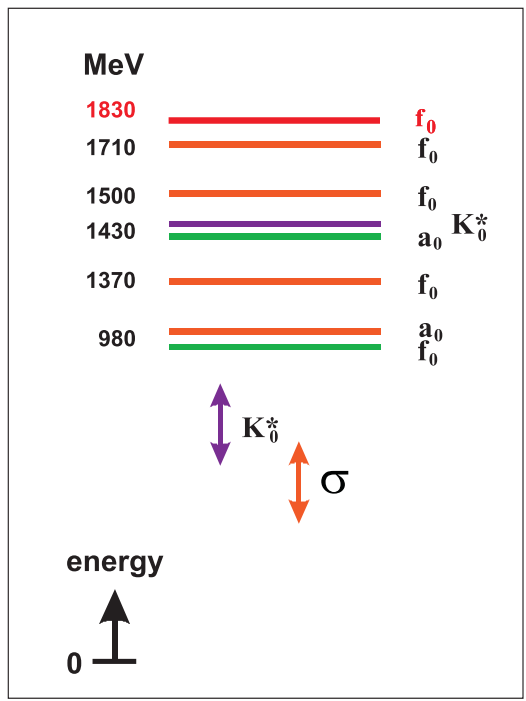}
\caption{Spectrum of $J^{PC}\,=\,0^{++}$ states in the Particle Data Tables~\cite{pdg}.} \label{fig:spectrum}
\end{center}
\vspace{-2.mm}
\end{figure}

We expect there to be a simple quark model $^3P_0$ multiplet. We have long known~\cite{morgan-respect} this cannot be readily identified with the nine lightest scalars in Fig.~1 ---
formed from the isotriplet $a_0$, an isodoublet $K^*_0$ and two $f_0$'s. The argument goes: how can the $a_0(980)$ and $f_0(980)$ be degenerate in mass with strong couplings to ${\overline K}K$, when a simple nonet would have just one  ${\overline s}s$ system? Jaffe~\cite{jaffe} has long predicted the existence of multi-quark mesons, which in the case of the scalars would  form a nonet with lower mass than the simple ${\overline q}q$ multiplet. 
Many have identified this tetraquark, ${\overline{qq}}qq$, system with the nine lightest scalars. In this nonet the $a_0, f_0(980)$ are the two heaviest states.
Both built from an $[sn]$ diquark and $[{\overline{sn}}]$ anti-diquark, with $n=u,d$, they are naturally degenerate in mass. In the non-relativistic potential model of
Weinstein and Isgur~\cite{wi}, however, it is only these tetraquark states with  their ${\overline K}K$ configurations that bind. The nonet would be reduced to a quartet. In either picture the heavier $a_0$ at 1450 MeV defines the isotriplet central line of the ${\overline q}q$ nonet. Of course, ${\overline q}q$ and tetraquark mesons are not orthogonal. They inevitably mix, through their common decay channels, like $\pi\pi$, ${\overline K}K$.

Almost by definition the simple quark model considers only the simplest components in a hadron's Fock space.
For the well-known vector and tensor mesons, like the $\rho^+$ and $a_2^+$, this is just $u{\overline d}$.  For, the $\phi$, it is ${\overline s}s$. Of course, the Fock space of each is really more complicated. The $\phi$ has ${\overline K}K$ components, the $\rho$ has $\pi\pi$, the $a_2$ has $\rho\pi$. It is through these that the $\phi$, $\rho$ and $a_2$, respectively, decay. Though present these \lq\lq multiquark'' components have a relatively small effect on the structure
of these mesons. The ${\overline q}q$ components dominate, and the picture of the simple quark model works.  In contrast for scalars, the channels to which they decay are crucial. While they, like the vectors and tensors,  have dominant two pseudoscalar decays, they have couplings that are not only intrinsically larger but $S$-wave. This means the  decay channels have a dramatic effect. With scalars seeded by ${\overline q}q$ configurations, new states are generated dynamically. The isotriplet and an isosinglet appear close to ${\overline K}K$ threshold as first found by van Beveren and collaborators~\cite{vanbeveren}. These then have sizeable ($\sim 40$\%) ${\overline K}K$ components in their wavefunctions~\cite{tornqvist}. The quark model works where decay channels are unimportant, but for scalars they are an intrinsic part of their make-up.

A further complication is that lattice calculations universally predict that even in a world without quarks there would be a spectrum of colorless hadrons made of glue. Of these the scalar is always the lightest~\cite{glueball-teper}. Predictions, by UKQCD~\cite{glueball-teper}, by Weingarten and by Morningstar and Peardon~\cite{glueball-mp}, range from 1500 to 1750 MeV. Experiment reveals states
at 1500 MeV largely explored by Crystal Barrel~\cite{amsler}, at 1710 MeV first found by Crystal Ball and confirmed by Mark III~\cite{mark3-glue} at SLAC (initially as a tensor, but then revised to be a scalar~\cite{pdg}), and most recently at 1830 MeV by BES as an $\omega\phi$ enhancement in $J/\psi$ radiative decay~\cite{bes-glue}. Each of these is claimed to be the glueball, or at least largely gluish. Each has their protagonists~\cite{weingarten,amsler-close,bicudo}. No argument is yet conclusive. Scalars with a significant mixture of glue may well exist, but which states have this mixture is still a matter of debate.

\section{$f_0(980)$: a ${\overline q}q$ or ${\overline sq}sq$ scalar or ${\overline K}K$ molecule?}

 Is the deuteron a six quark state, or is it a bound state of two baryons? If the distinction is that a six quark state is generated by short range inter-quark forces, while a bound state of a proton and a neutron is formed by longer range inter-hadron interactions, then Weinberg~\cite{weinberg} showed that this question can be answered.
 This is closely related 
to the issue of whether a CDD pole is needed in an $N/D$ description of the relevant scattering amplitudes~\cite{cdd}: an argument Dalitz~\cite{dalitz} applied to the nature of the $\Lambda(1405)$.
Morgan~\cite{morgan} showed that this same idea can in principle decide whether the $f_0$ and $a_0(980)$ mesons
are intrinsically  quark states or $K{\overline K}$ molecules~\cite{wi,achasov-kk}.
  Data from heavy meson decays are now accumulating significant statistics on channels like $\pi\pi$, $\pi\eta$ and ${\overline K}K$ where the $f_0$ and $a_0$ appear, with  sufficient precision and detail to address this.

The method relies on  established notions of $S$-matrix theory. 
 The partial waves of scattering and production amplitudes are analytic functions of c.m. energy. At each threshold the sheets of the energy plane bifurcate. A typical resonance has poles on a whole series of sheets.
It is the pole on the nearest unphysical sheet that normally controls experiment. The reflections on other sheets are further away and rather like a series of reflections in parallel mirrors the more distant have less and less impact. However, a Breit-Wigner resonance with a position close to a threshold to which it strongly couples (as exemplified by a Flatt\'e form~\cite{flatte}) will have poles on sheets that are equally nearby and {\bf both} will be required to shape the features seen in experiment. For the $f_0(980)$  the relevant sheets are four in number being defined by the $\pi\pi$ and ${\overline K}K$ channels. These sheets can be unfolded in the neighborhood of ${\overline K}K$ threshold, by considering the $k_2$-plane, where $k_2$ is the centre-of-mass 3-momentum in the ${\overline K}K$ channel, as shown in Fig.~2. Experiment is performed on the edge of sheet I, moving down the imaginary axis up to
$K^+K^-$ threshold, then round to ${\overline K^0}K^0$ threshold, and then along the real axis as the energy increases.  A resonance with a conventional Breit-Wigner shape, coupling largely to $\pi\pi$, would have poles on sheets II and III equidistant from the axes where experiment is performed.  Such a resonance is generated by short range forces.  Crudely speaking, each pole causes the phase-shift to rise by $\sim 90^o$~\cite{mp-kk}.
In contrast a state generated by longer range hadronic forces only has a pole on  the 2nd sheet~\cite{morgan}. If there were no coupling to $\pi\pi$, this would be a bound state with a pole on the imaginary axis. Its coupling to $\pi\pi$ moves this away into sheet II. In some rough sense, it is \lq\lq half'' a Breit-Wigner resonance.
 Precision  data on the $f_0$ and $a_0(980)$ in this region can in principle differentiate between the need for one pole or two~\cite{mp-kk,hanhart-kk}.

\begin{figure}[t]
\begin{center}
\includegraphics[width=7.7 cm]{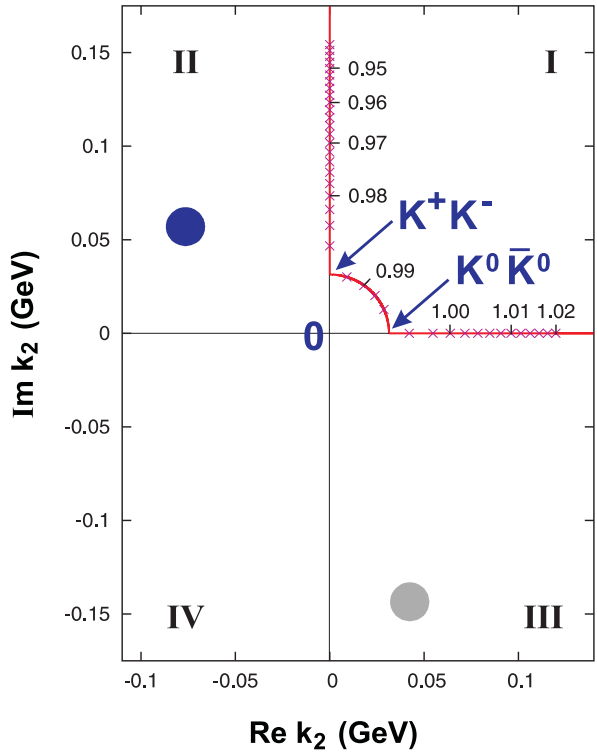}
\caption{The complex $k_2$-plane in the neighborhood of the two ${\overline K}K$ thresholds, where $k_2$ is the c.m. 3-momentum of the ${\overline K}K$ channel. The c.m. energy is marked ($\times$) every 20 MeV, with the energy in GeV enumerated every 100 MeV. The circles illustrate the position of poles on Sheets II and III.} \label{fig:k2}
\end{center}
\vspace{-5mm}
\end{figure}

Ever since its discovery, the $f_0(980)$ has been known to have a distorted shape, as seen in Fig.~3. Indeed, this is  well described by the Flatt\'e variation~\cite{flatte} of a Breit-Wigner form.
 By representing the $S$-matrix elements as quotients of Jost functions, the number of poles in the neighborhood of ${\overline K}K$ threshold can then be controlled~\cite{mp-kk}. 
The $\pi\pi$ phase-shift from the classic CERN-Munich experiment~\cite{CM}, the corresponding behavior of the inelasticity and the phase in $\pi\pi\to {\overline K}K$~\cite{etkin,mp-kk} allow equally good descriptions with both one or two nearby poles. However, these data are in rather wide bins. Fortunately, heavy flavor decays studied in $e^+e^-$ colliders add key information. We have
from Mark III~\cite{mark3-psi}, DM2~\cite{dm2-psi} and BES~\cite{bes-psi} results on
$J/\psi\to\phi(MM)$ decay with $M=\pi$ and $K$. Though the more recent BES data have the highest statistics, they have
 been binned in 30 MeV intervals. So it is the older results from Mark III in 10 MeV bins that 
are the most constraining. Nevertheless, they too allow both one pole and two pole descriptions of equal accuracy~\cite{pw}.

\begin{figure}[t]
\begin{center}
\includegraphics[width=12.0 cm]{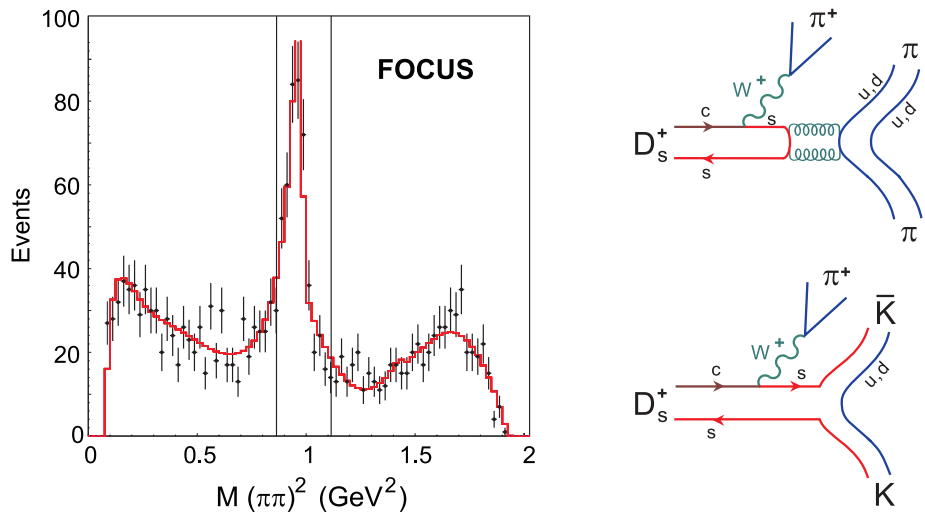}
\caption{$D_s^+\to (MM)\pi^+$ decay. On the left is an illustration of the $\pi^+\pi^-$ mass spectrum as observed by the FOCUS experiment~\cite{focus}
 --- the vertical lines around the $f_0(980)$ peak delineate the region employed in the Jost function analysis of Ref.~\cite{pw}. On the right are the corresponding quark line graphs appropriate to $\pi\pi$ and ${\overline K}K$ final states. } \label{fig:Ds}
\end{center}
\vspace{-7.mm}
\end{figure}

  What we need are precision results in small energy bites. Fortunately, these are provided by
$D_s\to\pi(MM)$ decays. To see the quality we show in Fig.~3, the $\pi\pi$ mass spectrum for this decay from the FOCUS collaboration~\cite{focus}. The region close to ${\overline K}K$ threshold used in this analysis is indicated by the 
vertical lines around the $f_0(980)$ peak in Fig.~3.
What makes the most difference are data from BaBar, as these have been partial wave analysed in both the $\pi^+\pi^-$ and $K^+K^-$ channels~\cite{marco}. While these results are {\it under wraps} and cannot be displayed, the ${\overline K}K$ data are in 4 MeV bins. When the $S$-wave is fitted together with the data on classic meson scattering experiments and $J/\psi$ decay, discussed above, these $D_s$ amplitudes
constrain the position of the poles rather well. Still fits
with both one and two nearby poles are equally likely. Seemingly we cannot distinguish between a molecule and an intrinsically quark state. The pole on Sheet II (indicated in Fig.~2) is at essentially the same location in both solutions. 
How can it be that fits that are so different in structure in the nearby complex plane can describe data nearly identically on the real axis? Since the presence of the Sheet III pole is the difference, exploration of the amplitudes reveals that the fit positions its Sheet III pole just where its residues
in both the $\pi\pi$ and ${\overline K}K$ channels are vanishingly small~\cite{pw}.
Thus, the two pole fit is really just pretending to be the single pole fit, and so a structure for the $f_0(980)$ generated by longer range inter-hadron forces
dominates. This would be exciting, if it were the definitive conclusion. However these are only preliminary conclusions~\cite{pw} of an analysis based on preliminary results from BaBar. Definitive  conclusions will have to wait till BaBar have finalised their partial wave analysis.

\section{Radiative width of the low mass scalars}

Another way to learn about the constitution of a hadron is to meaure its coupling to photons. For instance the coupling of vector mesons to $e^+e^-$ is readily predicted in the simple quark model, and experiment confirms this to be correct. For scalars (as with tensors) the coupling is to two photons, Fig.~4. The quark model predicts that the radiative width of each is determined by the square of the mean charge squared of their constituents, Fig.~4. For the tensors the ratio of radiative widths accords with the expectation for an ideally mixed quark model multiplet.  The corresponding predictions exist for the light spinless mesons. For a $({\overline u}u+{\overline d}d)$ scalar the two photon width is  5-10 keV at a mass of 700 MeV, reducing to 2-4 keV if the mass is down at 500 MeV~\cite{chano}, while for an ${\overline s}s$ state it is predicted to be $\sim 0.2-0.4$ keV~\cite{barnes,anisovich}. Calculations for a ${\overline K}K$ molecule are $\sim 0.2-0.6$ keV~\cite{achasov-gg,barnesKK,hanhart-gg}.  However one should treat these with some care. 
\begin{figure}[bh]
\begin{center}
\includegraphics[width=12.5 cm]{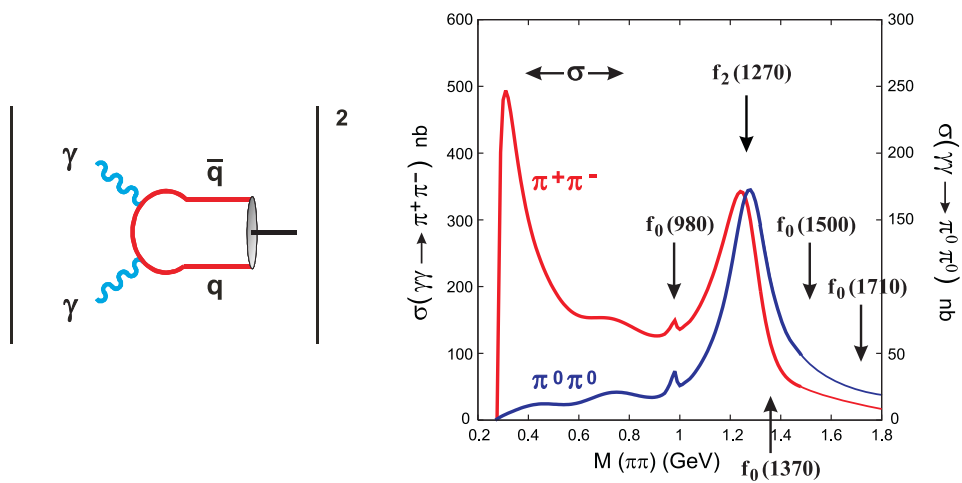}
\caption{Two photon process: on the left a graph depicting the radiative width of a ${\overline q}q$ bound state. On the right is a sketch of the integrated cross-section for $\gamma\gamma\to\pi^+\pi^-$ and $\pi^0\pi^0$ with different scales. Below 1.5 GeV this sketch is based on data from Mark II~\cite{mark2-gg}, CELLO~\cite{cello-gg}, Belle~\cite{belle-gg} and Crystal Ball~\cite{cb-gg}. The positions of key resonances are indicated. There is as yet no evidence for the sharp peaking near 1~GeV in the $\pi^0\pi^0$ channel, or for any structures above 1.5 GeV to associate with the higher scalars. } \label{fig:twophoton}
\end{center}
\vspace{-4.mm}
\end{figure}
Experience with the pseudoscalars $\pi^0$, $\eta$ and $\eta'$, 
shows that  agreement with experiment only comes if the quark model prediction is assumed to be bolted on to a  factor of ${mass}^3$~\cite{haynes}. That such an {\it ad hoc} correction is required illustrates that for light hadrons, including the scalars, a truly relativistic strong coupling calculation is essential.

How do we extract the two photon widths from experiment? The two channels $\gamma\gamma\to\pi^+\pi^-$ and $\pi^0\pi^0$ are those with the most data. The charged and neutral pion results are starkly different close to threshold as sketched in
Fig.~4. The charged channel is dominated by the one-pion exchange Born term, while that in the neutral case is very small (a factor of 40 or 50 less). It might appear that it is in the $\pi^0\pi^0$ channel that one should look for the $\sigma$ and extract its two photon coupling. Simply given a cross-section of $\sim 10$~nb, this might suggest a glueball character. This is the idea of the {\it red dragon} of Minkowski and Ochs~\cite{minkochs}, but for them this scalar is at higher mass away from threshold. In contrast, the $\sigma$ with a pole at $E = 441 -i272$ MeV~\cite{ccl} is crouching at much lower mass: see Fig.~5. Its two photon coupling is controlled by final state interactions. 
 Despite the neutral cross-section being small the \lq\lq effective''
 radiative width of the $\sigma$
is not $\sim 0.1$ keV but several keV~\cite{mp-prl} :
  telling us that quarks are an essential part of the $\sigma$'s make-up.
\begin{figure}[h]
\vspace{3.mm}
\begin{center}
\includegraphics[width=12. cm]{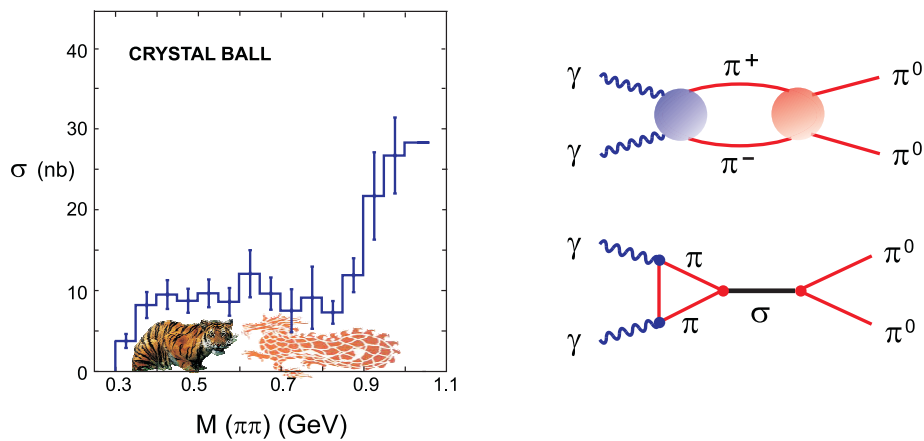}
\caption{Cross-section for $\gamma\gamma\to\pi^0\pi^0$, integrated over the angular range $|\cos\theta^*| \le 0.8$ from Crystal Ball~\cite{cb-gg}. The dominant mechanism for generating the $\pi^0\pi^0$ final state is through the process $\gamma\gamma\to\pi^+\pi^-$ by  one-pion-exchange, followed by $\pi^+\pi^-\to\pi^0\pi^0$: the $\sigma$ is one component of this, as shown in the graphs on the right. Though the $\sigma$ has an effective two photon width of several keV, the final state interactions in this channel mean that it  is hidden, as indicated by the \lq\lq crouching tiger''. 
 At slightly higher mass Minkowski and Ochs~\cite{minkochs} have claimed a scalar -- the {\it red dragon} -- is there. } \label{fig:tiger}
\end{center}
\end{figure}

 Now let us turn to the $f_0(980)$. Though earlier results from Mark II~\cite{mark2-gg} and CELLO~\cite{cello-gg} hinted at a shoulder in the $\gamma\gamma\to\pi^+\pi^-$ cross-section around 1~GeV, Belle  with data in 5 MeV bins in $\pi\pi$ mass is the first experiment to reveal a clear peak, sketched in Fig.~4. However, extracting the radiative width for this narrow state is still not easy. Belle fit their integrated cross-section and find 
$\Gamma(f_0(980)\to\gamma\gamma)\,=\,(0.205\pm 0.089 \pm 0.132)$ keV~\cite{belle-gg}. The errors reflect the marked systematic sensitivity to the assumed non-resonant background and the presumption of pure helicity two for the $f_2(1270)$.   The way to proceed is to perform an Amplitude Analysis using all the available data on charged and neutral channels and all the angular information to separate out the $I=0$ $S$-wave signal. This is non-trivial given that the charged pion data cover no more than 60\% of the angular range and the contributions of individual partial waves are then not orthogonal. Preliminary results of the Amplitude Analysis from the Belle group and myself~\cite{belle-mp} show that solutions with a radiative width for the $f_0(980)$ from  0.1 to 0.4 keV are equally possible. These solutions differ in the amount $S$-wave and helicity zero $D$-wave.  This is not better determined because
Belle has difficulty in separating the background $\mu^+\mu^-$ pairs. This results in a rather poor angular distribution in the key  0.80 and 1.05 GeV region. Finer $\pi^0\pi^0$ differential cross-sections covering a bigger angular range, perhaps 80\%, are promised soon from Belle and these will hopefully allow a cleaner amplitude separation and a more precise width for the $f_0(980)$. Exactly how many tenths of a keV will be critical in confirming its sizeable ${\overline K}K$ composition.
Once this is settled we can go on and study the higher mass scalar states, like the $f_0(1370)$, $f_0(1500)$ and $f_0(1710)$ as indicated in Fig.~4. That is for the future.

\section{Conclusions}

The structure of the scalars is far from simple. The isoscalar members couple to the vacuum of QCD  and reflect its structure. The fact that there is not just one scalar, but many, may be echoed as GeV scales give way to TeV and we learn, as the LHC has its first collisions, about the scalar(s) that break electroweak symmetry.

\section*{Acknowledgments}
It is a pleasure to thank the organisers
 for the invitation to give this talk.
The work presented here was partially supported by the EU-RTN Programme, 
Contract No. MRTN-CT-2006-035482, \lq\lq FLAVIAnet''. 


\end{document}